\documentclass[nofootinbib,prd,twocolumn,showpacs,showkeys,preprintnumbers]{revtex4-1}
\usepackage{hyperref,amssymb,amsmath,mathrsfs,bm,graphicx}
\begin{document}
\title {The Evolution of Self-Gravitating Radiating Geodesic Fluid Spheres Instantaneously Admitting a Time-Like Killing}
\author{L. Herrera}
\email{lherrera@usal.es}
\affiliation{Instituto Universitario de F\'isica
Fundamental y Matem\'aticas, Universidad de Salamanca, Salamanca 37007, Spain.}
\author{A. Di Prisco}
\email{alicia.diprisco@ucv.ve}
\affiliation{Escuela de F\'{\i}sica, Facultad de Ciencias,
Universidad Central de Venezuela, Caracas 1050, Venezuela.}
\begin{abstract}
In this study we analyze the effects produced by the sudden appearance  of a time-like Killing vector (TKV),  on  the evolution of   radiating geodesic fluids. This is achieved by   resorting  to the concept of asymmetry factor which  allows the instantaneous appearance  of a given symmetry  for a given value of the time-like coordinate.  All relevant physical variables are analyzed in detail showing the imprints of the emergence of TKV at some point of the evolution. The main lesson that can be extracted from the obtained results is that the sudden appearance of a TKV, even for an infinitesimal  period of time, drastically changes the subsequent evolution of the system. Of particular relevance is the appearance of a thermal effect linking  the asymmetry factor and the relaxation time. The potential applications  of these  results, on the study of gravitational collapse, are  discussed at the end of the paper.
 \end{abstract}
\date{\today}
\maketitle

{\it To our friend Lluis Bel, In memoriam}

\section{Introduction}

In a recent paper~\cite{1}, we introduced the concept of the asymmetry factor, which is a scalar function allowing one %MDPI EE: Please check intended meaning has been retained.Ok
 to switch on (or off) a given symmetry for any value  of the time-like coordinate. More specifically, in~\cite{1}, we considered the case of a conformal Killing vector, and~we looked for the imprints that the instantaneous appearance of such  a symmetry produces on the evolution of the fluid. In~this work, we want to apply a similar scheme to a case where the involved symmetry  corresponds to a time-like Killing vector field, and~the matter content corresponds to  geodesic dissipative  fluids. This type of fluid has been shown to be very useful in the study of gravitational collapse (see~\cite{gr1a,gr1,gr2,gr3,gr4,gr5,gr6} and \linebreak the references therein).

The main reason to consider a TKV as the instantaneously appearing symmetry is that radiating spherically symmetric fluid distributions, in general, do not admit a TKV; therefore, we wonder what influence this might have on the evolution of the system. %MDPI EE: Please check intended meaning has been retained. See the change.
 In~other words, we wanted to find out how to describe the evolution of the system in terms of the factor measuring deviations from the time invariance (the asymmetry factor). 

Thus, we shall assume geodesic spherically symmetric fluid distributions that, at some point of their  evolution, admit a TKV. %MDPI EE: Please check intended meaning has been retained. Ok
 The~(dis)appearance of the TKV is controlled by a tensor defined through the asymmetry factor. As~we will see below, the sudden appearance of the TKV, even during  an infinitesimal small interval of time, has a remarkable influence on all physical~variables.

All relevant variables characterizing the obtained model are  analyzed in detail, and~the  influence of the appearance of the TKV on the behavior of the complexity factor, the~anisotropic factor, the~luminosity, the~temperature and   the redshift is~emphasized.

The method and the variables required for describing  the evolution of  spherically symmetric fluid distributions  are  presented  in the next section.  In Section 3, %MDPI: We revised the section citation format, please cnfirm. The same below. Ok
we introduce the concept of an asymmetry factor for a time-like isometry, and~express the metric variables in terms of such a factor.
A specific model  is analyzed in Section 4, and~the obtained results are discussed in Section 5. %MDPI: We revised the Section number, please cnfirm. Ok
 Finally, Appendix A is~included.

\section{Fundamental variables  and relevant equations}
In this section we introduce all variables and definitions allowing for a full description of the structure and evolution of geodesic, spherically symmetric dissipative fluids. For detail see  \cite{ps2}.

For  spherically symmetric distributions  of evolving
geodesic fluids, bounded  by a spherical surface $\Sigma$, the general interior metric in comoving coordinates reads
\begin{equation}
ds^2=-dt^2+B^2dr^2+R^2(d\theta^2+\sin^2\theta d\phi^2),
\label{1}
\end{equation}
where  $B$ and $R$ are functions of $t$ and $r$ and are assumed
positive.  Observe that  $B$ is dimensionless, whereas $R$ has the same dimension as $r$.

The matter content consists in 
an anisotropic fluid (principal stresses unequal) undergoing dissipation in the
form of heat flow (diffusion approximation).

Thus, the energy--momentum tensor $T_{\alpha\beta}$ may be written as , 
\begin{eqnarray}
T_{\alpha\beta}&=&(\mu +
P_{\perp})V_{\alpha}V_{\beta}+P_{\perp}g_{\alpha\beta}+(P_r-P_{\perp})K_{
\alpha}K_{\beta}\nonumber \\&+&q_{\alpha}V_{\beta}+V_{\alpha}q_{\beta}
, \label{3}
\end{eqnarray}
or
\begin{equation}
T_{\alpha \beta} = {\mu} V_\alpha V_\beta + P h_{\alpha \beta} + \Pi_{\alpha \beta} +
q \left(V_\alpha K_\beta + K_\alpha V_\beta\right), \label{Tab}
\end{equation}
where $\mu$ is the energy density, $P_r$ the radial pressure,
$P_{\perp}$ the tangential pressure, $q^{\alpha}$ the heat flux, $V^{\alpha}$ the four-velocity of the fluid,
and $K^{\alpha}$ a unit four-vector along the radial direction, and 
\begin{eqnarray}
V^{\alpha}V_{\alpha}=-1, \;\; V^{\alpha}q_{\alpha}=0, \;\; K^{\alpha}K_{\alpha}=1,\;\;
K^{\alpha}V_{\alpha}=0.
\end{eqnarray}

$$ P=\frac{P_{r}+2P_{\bot}}{3}, \qquad h_{\alpha \beta}=g_{\alpha \beta}+V_\alpha V_\beta,$$

$$\Pi_{\alpha \beta}=\Pi\left(K_\alpha K_\beta - \frac{1}{3} h_{\alpha \beta}\right), \qquad \Pi=P_{r}-P_{\bot}.$$

\begin{eqnarray}
V^{\alpha}&=&\delta_0^{\alpha}, \;\;
q^{\alpha}=qK^{\alpha}, \;\;
K^{\alpha}=B^{-1}\delta^{\alpha}_1,
\end{eqnarray}
where $q$ is a function of $t$ and $r$.

The Einstein equations for (\ref{1}) and (\ref{Tab}), are explicitly written  in Appendix A.

The expansion $\Theta$ of the fluid is
given by
\begin{equation}
\Theta={V^{\alpha}}_{;\alpha}, \label{4b}
\end{equation}
and its  shear $\sigma_{\alpha\beta}$ by
\begin{equation}
\sigma_{\alpha\beta}=V_{(\alpha
;\beta)}+a_{(\alpha}V_{\beta)}-\frac{1}{3}\Theta h_{\alpha\beta}.
\label{4a}
\end{equation}

From  (\ref{4b}) we have  for the expansion
\begin{equation}
\Theta=\left(\frac{\dot{B}}{B}+2\frac{\dot{R}}{R}\right),
\label{5c1}
\end{equation}
where the  prime stands for $r$
differentiation and the dot stands for differentiation with respect to $t$.

The shear tensor (\ref{4a}) has only one independent non zero component, and may be written as 
\begin{equation}
\sigma_{\alpha \beta}=\sigma \left(K_\alpha K_\beta-\frac{h_{\alpha \beta}}{3}\right),
\label{defs}
\end{equation}
with 
\begin{equation}
\sigma=\left(\frac{\dot{B}}{B}-\frac{\dot{R}}{R}\right).\label{5b1}
\end{equation}

Next, the mass function $m(t,r)$  reads \cite{ms,kt}
\begin{equation}
m=\frac{R^3}{2}{R_{23}}^{23}
=\frac{R}{2}\left[\dot R^2-\left(\frac{R^{\prime}}{B}\right)^2+1\right].
 \label{17masa}
\end{equation}

Introducing the proper time derivative $D_T$
given by
\begin{equation}
D_T=\frac{\partial}{\partial t}, \label{16}
\end{equation}
we can define the areal velocity velocity $U$  as the variation of the areal radius with respect to proper time, i.e.
\begin{equation}
U=D_TR , \label{19}
\end{equation}
where $R$ defines the areal radius of a spherical surface inside the fluid distribution (as
measured from its area).

Then (\ref{17masa}) can be rewritten as
\begin{equation}
E \equiv \frac{R^{\prime}}{B}=\left(1+U^2-\frac{2m}{R}\right)^{1/2}.
\label{20x}
\end{equation}
Using (\ref{20x}) we can express (\ref{17a}) as
\begin{equation}
4\pi q=E\left[\frac{1}{3}D_R(\Theta-\sigma)
-\frac{\sigma}{R}\right],\label{21a}
\end{equation}
where   $D_R$ denotes the proper radial derivative,
\begin{equation}
D_R=\frac{1}{R^{\prime}}\frac{\partial}{\partial r}.\label{23a}
\end{equation}

\subsection{The complexity factor}
The idea of complexity in different branches of science  has a long and a venerable history (see \cite{ts} and references therein). Some years ago important advances on  this issue were achieved by L\'opez-Ruiz and collaborators within  the context of statistical physics \cite{3,3bisbis,4,4bis}. This approach was latter on employed by some authors to define complexity for self-gravitating systems (see \cite{10}  and references therein). However such a definition suffers from some flaws, which  motivated the search for an alternative definition of complexity for self-gravitating systems. This search led to the definition proposed in \cite{ps1,ps2}, which is based on the concept of complexity factor.

This is a  scalar function measuring  the degree of complexity of a fluid distribution \cite{ps1,ps2}. Mathematically it is the  scalar defining the trace-free part of the electric Riemann tensor ($Y_{TF}$)  \cite{sc}.

Thus, the tensor $Y_{\alpha \beta}$  defining the electric part of the Riemann tensor   given  by 
\begin{equation}
Y_{\alpha \beta}=R_{\alpha \gamma \beta \delta}V^\gamma V^\delta,
\label{electric}
\end{equation}
may be written as 

\begin{eqnarray}
Y_{\alpha\beta}=\frac{1}{3}Y_T h_{\alpha
\beta}+Y_{TF}(K_{\alpha} K_{\beta}-\frac{1}{3}h_{\alpha
\beta})\label{electric'}.
\end{eqnarray}

The scalar $Y_{TF}$  is identified with the complexity factor and may be written  in terms of the anisotropy of pressure, the density inhomogeneity and  the dissipative variables as (see \cite{sc, ps1, ps2} for details)
\begin{equation}
Y_{TF}= -8\pi\Pi +\frac{4\pi}{R^3}\int^r_0{R^3\left(D_R {\mu}-3{q}\frac{U}{RE}\right)R^\prime dr}.
\label{Y}
\end{equation}

In terms of the metric functions the scalar $Y_{TF}$ reads

\begin{eqnarray}
Y_{TF}= \frac{\ddot R}{R} - \frac{\ddot B}{B} .
\label{itfm}
\end{eqnarray}

\subsection{The exterior spacetime and junction conditions}
For a smooth matching of (\ref{1}) with Vaidya on the boundary surface of the fluid distribution  we have to assume the Darmois conditions  \cite{14nn}

Thus, outside $\Sigma$ we assume that we have the Vaidya
spacetime 
described by
\begin{equation}
ds^2=-\left[1-\frac{2M(v)}{\rho}\right]dv^2-2d\rho dv+\rho^2(d\theta^2
+\sin^2\theta
d\phi^2) \label{1int},
\end{equation}
where $M(v)$  denotes the total mass,
 $v$ is the retarded time, and $\rho$ is a null coordinate.

The matching of the full non--adiabatic sphere  (including viscosity) to
the Vaidya spacetime, on the surface $r=r_{\Sigma}=$ constant, was discussed in
\cite{chan1}. It implies:
\begin{itemize}
\item From the continuity of the first  differential form, 
\begin{equation}
dt\stackrel{\Sigma}{=}dv \left(1-\frac{2M(v)}{\rho}\right)\stackrel{\Sigma}{=}d\tau, \label{junction1f}
\end{equation}
\begin{equation}
R\stackrel{\Sigma}{=}\rho(v), \label{junction1f2}
\end{equation}
and 
 \begin{equation}
\left(\frac{dv}{d\tau}\right)^{-2}\stackrel{\Sigma}{=}\left(1-\frac{2m}{\rho}+2\frac{d\rho}{dv}\right), \label{junction1f3}
\end{equation}
where $\tau$ denotes the proper time measured on $\Sigma$, and  $\stackrel{\Sigma}{=}$ means that both sides of the equation
are evaluated on $\Sigma$.

\item From the continuity of the second differential form
\begin{equation}
m(t,r)\stackrel{\Sigma}{=}M(v), \label{junction1}
\end{equation}

\begin{equation}
q\stackrel{\Sigma}{=}P_r.\label{j3}
\end{equation}
\end{itemize}

Also, we have
\begin{equation}
q\stackrel{\Sigma}{=}\frac{L}{4\pi \rho^2}, \label{20lum}
\end{equation}
where $L_\Sigma$ denotes   the total luminosity of the  sphere as measured on its surface and is given by
\begin{equation}
L \stackrel{\Sigma}{=}L_{\infty}\left(1-\frac{2m}{\rho}+2\frac{d\rho}{dv}\right)^{-1}, \label{14a}
\end{equation}
and where
\begin{equation}
L_{\infty} =-\frac{dM}{dv}\stackrel{\Sigma}{=} -\left[\frac{dm}{dt}\frac{dt}{d\tau}\left(\frac{dv}{d\tau}\right)^{-1}\right],\label{14b}
\end{equation}
is the total luminosity measured by an observer at rest at infinity, which is defined by 
\begin{equation}
L_\infty=4\pi P_{r\Sigma}\left[R^2\left(\frac{R^\prime}{B}+\dot R \right)^2 \right]_\Sigma.
\label{lum}
\end{equation}

The boundary redshift $z_\Sigma$ is given by
\begin{equation}
\frac{dv}{d\tau}\stackrel{\Sigma}{=}1+z,
\label{15b}
\end{equation}
with
\begin{equation}
\frac{dv}{d\tau}\stackrel{\Sigma}{=}\left(\frac{R^{\prime}}{B}+\dot R \right)^{-1}.
\label{16b}
\end{equation}

When Darmois conditions \cite{14nn} are not satisfied, the boundary surface is a thin shell and we have to assume the Israel conditions \cite{isjc}.

\subsection{The transport equation}
For obtaining an expression for the temperature we shall resort to the  transport equation proposed in the context of the Israel-- Stewart theory \cite{19nt, 20nt, 21nt} , which for a geodesic fluid reads
\begin{eqnarray}
\tilde \tau h^{\alpha \beta}V^\gamma q_{\beta;\gamma}+q^\alpha=-\kappa h^{\alpha \beta} T_{,\beta} -\frac{1}{2}\kappa T^2 \left(\frac{\tilde \tau V^\beta}{\kappa T^2}\right)_{;\beta} q^\alpha,
\label{tre}
\end{eqnarray}
where $\kappa$ denotes the thermal conductivity, and $T$ and $\tilde \tau$ denote temperature and relaxation time respectively. We recall that the  relaxation time is the time taken by the system to return  to the steady state after it has been suddenly removed from it.

In the spherically symmetric case under consideration, the transport equation has only one independent component which may be obtained from (\ref{tre}) by contracting with the unit spacelike vector $K^\alpha$, we get
\begin{equation}
\tilde \tau V^\alpha  q_{,\alpha}+q=-\kappa K^\alpha T_{,\alpha}-\frac{1}{2}\kappa T^2\left(\frac{\tilde \tau V^\alpha}{\kappa T^2}\right)_{;\alpha} q.
\label{5}
\end{equation}

In some cases (when the last term in (\ref{tre}) may be neglected \cite{t8}) it is possible to simplify the transport  equation, producing 
\begin{equation}
\tilde \tau V^\alpha  q_{,\alpha}+q=-\kappa K^\alpha T_{,\alpha}.
\label{5trun}
\end{equation}

\section{The asymmetry factor for isometries}
Spacetimes  admitting a TKV, satisfy  the equation
\begin{equation}
\mathcal{L}_\chi g_{\alpha \beta} =0, 
\label{1cmh}
\end{equation}
 where $\mathcal{L}_\chi$ denotes the Lie derivative with respect to the vector field $\chi$, which in our case is time-like, and  has the form
\begin{equation}
\chi= \partial _{\mathbf{t}}.
\label{2cmh}
\end{equation}

 In this work we shall consider fluid distributions which in general do not admit  a TKV, except at a given moment of their evolution.
 
 Thus let us consider 
\begin{equation}
\mathcal{L}_\chi g_{\alpha \beta}=H_{\alpha \beta},
\label{mod1n}
\end{equation}
implying
\begin{equation}
\chi^\rho \partial_\rho g_{\alpha \beta} +g_{\alpha \rho}\partial_\beta \chi^\rho+g_{\beta \rho}\partial_\alpha \chi^\rho=H_{\alpha \beta},
\label{mod11nn}
\end{equation}
where $H_{\alpha \beta}$ is (so far) an arbitrary symmetric tensor. Obviously  the vector $\chi$ defines a TKV  iff  $H_{\alpha \beta}=0$.

Using (\ref{2cmh}) one obtains for (\ref{1})
$H_{00}=H_{01}=H_{02}= H_{03}=H_{13}=H_{12}=  H_{23}=0$.

Next, let us  write the tensor $H_{\alpha \beta}$ in the form 

\begin{equation}
H_{\alpha \beta}=H_{T} \frac{g_{\alpha \beta}}{4}+H_{TF}  \left(K_\alpha K_\beta-\frac{1}{3}h_{\alpha \beta}\right),
\label{tkv1}
\end{equation}
where $H_T $ and $H_{TF}$ are two  dimensionless  scalar functions depending only on $t$.

Then, we obtain from the above
\begin{equation}
H_T=0,
\label{nda}
\end{equation}
\begin{equation}
\dot B=\frac{\alpha H_{TF} B}{3},
\label{nd2b}
\end{equation}

\begin{equation}
\dot R=-\frac{\alpha H_{TF} R}{6},
\label{nd3b}
\end{equation}
where $\alpha$ is a unit constant with dimensions $\frac{1}{length}$.

The scalar function $H_{TF}$ is the asymmetry factor. It enables us to switch on (off) the isometry at any given point of the evolution.

Using the above result we may write the metric functions as 
\begin{equation}
A=1,\quad B=g(r)e^{\frac{\alpha}{3}\int H_{TF}dt},\quad
\alpha R=h(r)e^{-\frac{\alpha}{6}\int H_{TF}dt},
\label{nd12b}
\end{equation}
where $g$ and $h$ are arbitrary dimensionless functions.

Using these  results  we shall build up  our  model.

At this point we should notice that  if  we assume  the  fluid to be  non-dissipative then feeding back (\ref{nd12b}) into 
 (\ref{13})  we see that $q\neq0$.

Thus, the non--dissipative case cannot be analyzed by means of the asymmetry factor scheme for a TKV and geodesic fluids.

\section{The model} 
Based on the above comments we are constrained  to  the dissipative case exclusively. 

Using the form (\ref{tkv1}) of the tensor $H_{\alpha \beta}$, and assuming for $H_{TF} $
\begin{equation}
H_{TF} =\alpha(t-t_0),
\label{nref}
\end{equation}
 we obtain 
\begin{equation}
A=1\quad B=g(r)e^{y^2/6}\quad
\alpha R=h(r)e^{-y^2/12},
\label{nd12}
\end{equation}
where $y\equiv\alpha(t-t_0)=H_{TF}$.

In this case any possible model is determined up to two functions of $r$.

In order to specify our model we shall proceed as follows.

Let us introduce the dimensionless  variable
\begin{equation}
x\equiv\frac{r}{r_\Sigma},
\label{mond1}
\end{equation}
and choose for the free parameters the value
\begin{equation}
 \alpha r_\Sigma=1.
\label{mond2}
\end{equation}

Based on the above,  the physical variables read
\begin{eqnarray}
8\pi \mu/\alpha^2 &=&-\frac{y^2}{12}-\left[\frac{2 h_{,xx}}{h}+\left(\frac{h_{,x}}{h}\right)^2-\frac{2g_{,x} h_{x}}{g h}\right] \frac{e^{-y^2/3}}{g^2}\nonumber \\&+&\frac{e^{y^2/6}}{h^2}
,\label{muidscosf}
\end{eqnarray}

\begin{eqnarray}
8\pi q/\alpha^2=-\frac{e^{-y^2/6}y h_{,x}}{gh},
\label{qdisb}
\end{eqnarray}

\begin{eqnarray}
8\pi P_r/\alpha^2=\frac{1}{3}-\frac{y^2}{12}+\frac{e^{-y^2/3}h_{,x}^2}{g^2 h^2}-\frac{e^{y^2/6}}{h^2},
\label{pridscosfb}
\end{eqnarray}

\begin{equation}
8\pi P_\bot/\alpha^2=-\frac{1}{6}\left(1+\frac{y^2}{2}\right)+\frac{e^{-y^2/3}}{g^2}\left(\frac{h_{,xx}}{h}-\frac{h_{,x}g_{,x}}{hg}\right),
\label{ptdscosfb}
\end{equation}

Also,  from (\ref{14a}) and (\ref{lum}) we obtain 
\begin{equation}
L_\Sigma\stackrel{\Sigma}{=}-\frac{yh h_{,x}e^{-y^2/3}}{2g},
\label{lums}
\end{equation}
and
\begin{eqnarray}
L_\infty \stackrel{\Sigma}{=}L_\Sigma e^{-y^2/2}\left(\frac{h_{,x}}{g}-\frac{y h e^{y^2/6}}{6}\right)^2,
\label{lumsin} 
\end{eqnarray}
whereas the expressions for the complexity factor and the redshift read
\begin{equation}
Y_{TF}/\alpha^2=-\frac{1}{2}\left(\frac{y^2}{6}+1\right),
\label{bhf}
\end{equation}
and 
\begin{equation}
z\stackrel{\Sigma}{=}\frac{6 ge^{y^2/4}}{6h_{,x}-hyg e^{y^2/6}}-1.
\label{zdis}
\end{equation}
Finally, using  the truncated version of the transport equation (\ref{5trun}), and (\ref{qdisb}) we may obtain an expression for the temperature. It reads
\begin{eqnarray}
\tilde T=\frac{1}{8\pi}\left[y-\tau^\ast\left(\frac{y^2}{3}-1\right)\right]\int{\frac{h_{,x}}{h}dx}  +\tilde T_0(y),
\label{tdis}
\end{eqnarray}
where $\tau^\ast=\alpha \tilde \tau$  and $\tilde T=\frac{\kappa T}{\alpha}$.
\begin{figure}[h]
\includegraphics[scale=0.7]{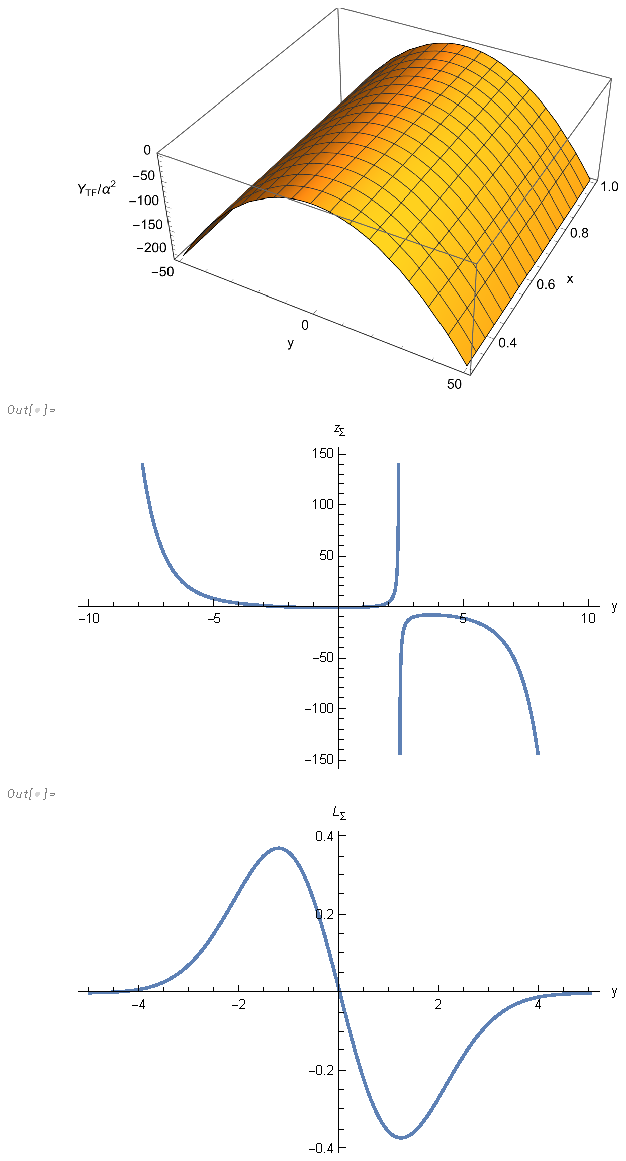}
\caption {$Y_{TF}/\alpha^2$  as function of $y$ and $x$  in the interval $(y,-50, 50)$, $(x, .3, 1)$;  $z_\Sigma$ in the interval $(y, -10, 10)$; $L_{\Sigma}$ in the interval $(y, -5, 5)$.}\label{Figure 1}
\end{figure}

\begin{figure}[h]
\includegraphics[width=5.in,height=6.in,angle=0]{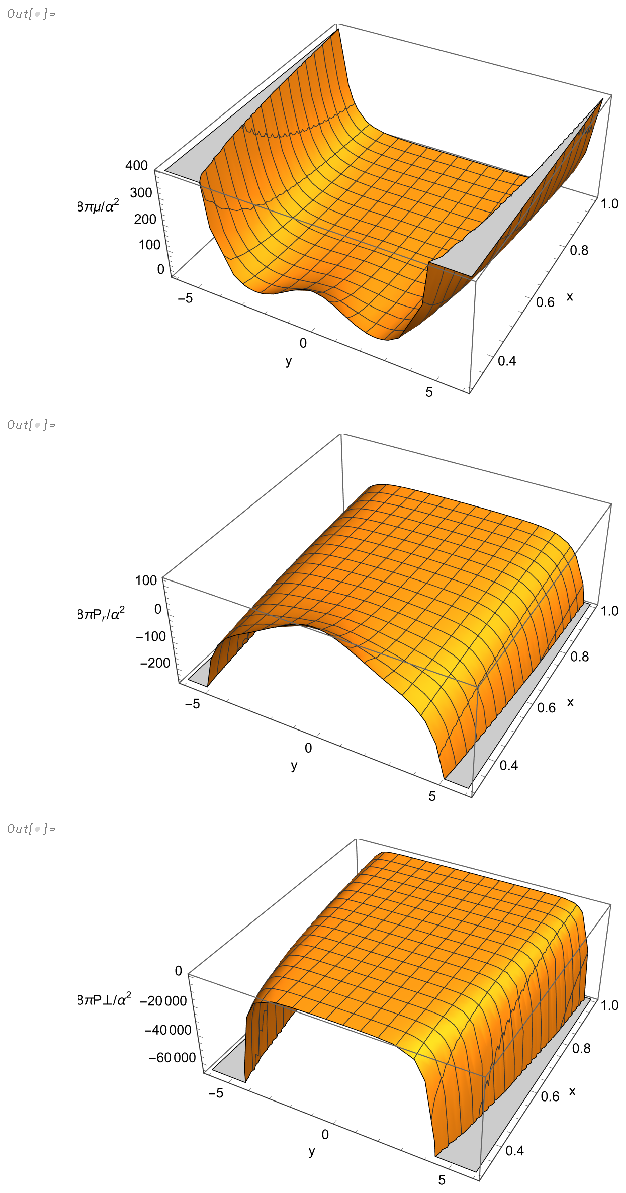}
\caption{$8\pi \mu/\alpha^2$, $8\pi P_r/\alpha^2$, $8\pi P_\bot/\alpha^2$ as functions of $y$ and $x$  in the interval $(y, -6, 6)$, $(x, .3,  1)$}.\label{Figure 2}
\end{figure}
.
\begin{figure}[h]
\includegraphics[width=5.in,height=6.in,angle=0]{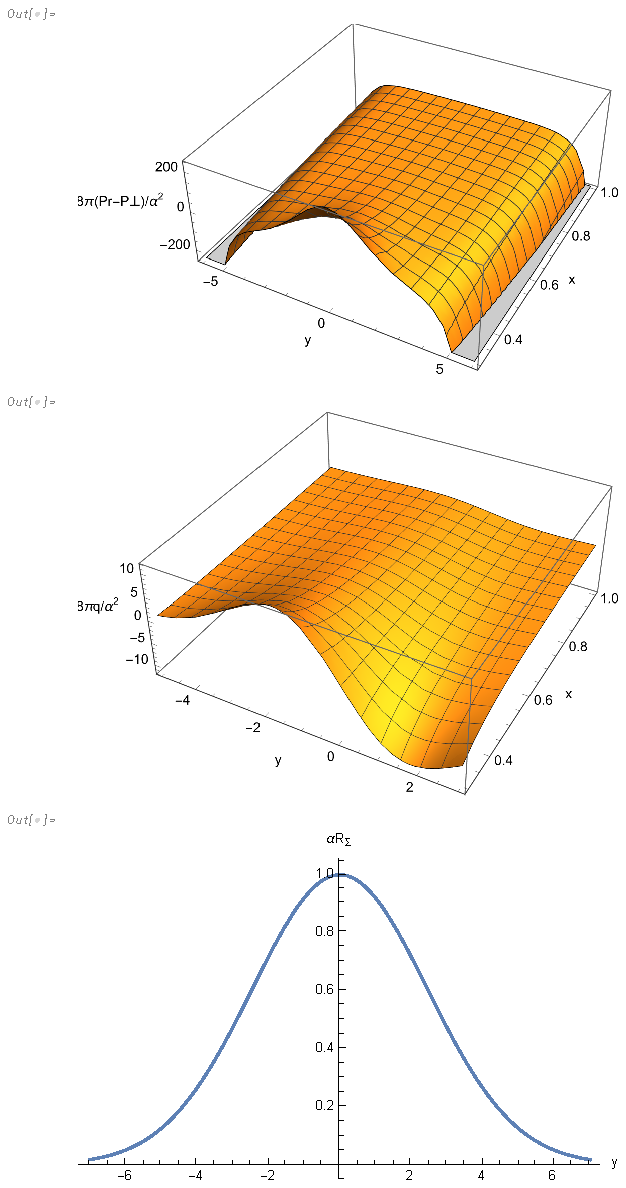}
\caption{$8\pi(P_r-P_\bot)/\alpha^2$ as function of $y$ and $x$ in the interval $(y, -6,  6)$, $(x, .3, 1)$; $8\pi q /\alpha ^2$  as function of $x$ and  $y$ in the interval $(y, -5, 3)$, $(x, .3, 1)$ and $\alpha R_\Sigma$ in the interval $(y, -7, 7)$}.\label{Figure 3}
\end{figure}

\begin{figure}[h]
\includegraphics[width=5.in,height=6.in,angle=0]{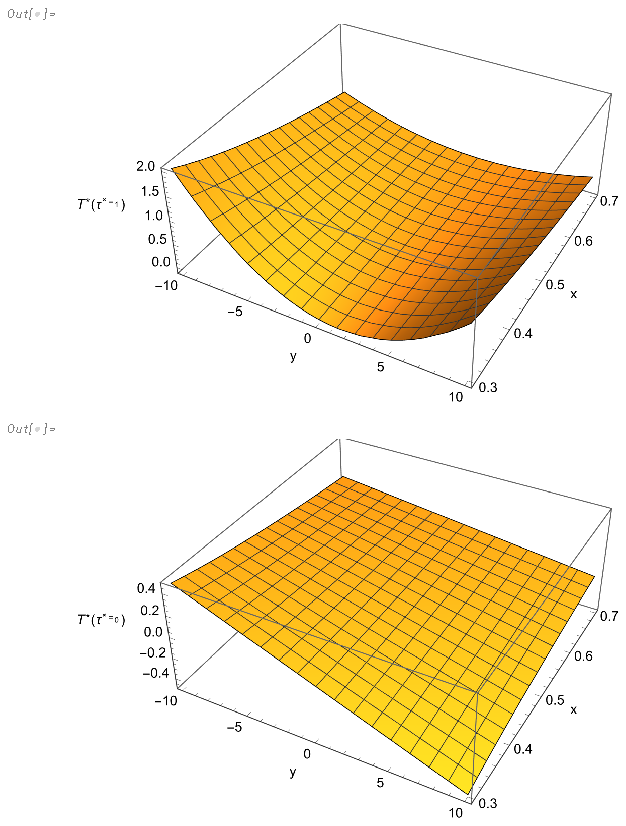}
\caption{$T^\ast$ as function of $y$ and $x$  in the interval $(y, -10, 10)$, $(x, .3,  .7)$ for $\tau^\ast=1$ and   $\tau^\ast=0$}.\label{Figure 4}
\end{figure}

For completely determine the model we still need to specify the functions $h$ and  $g$. 

For simplicity we choose
\begin{equation}
g(r)=h(r)=\alpha r=x.
\label{deff}
\end{equation}

From the above restrictions the physical variables become
\begin{eqnarray}
8\pi \mu/\alpha^2 =-\frac{y^2}{12}+\left(\frac{1}{x^2} +e^{y^2/2}\right)\frac{e^{-y^2/3}}{x^2},
\label{muidscosfa}
\end{eqnarray}

\begin{eqnarray}
8\pi q/\alpha^2=-\frac{ye^{-y^2/6}}{x^2},
\label{qdisb}
\end{eqnarray}

\begin{eqnarray}
8\pi P_r/\alpha^2=\frac{1}{3}-\frac{y^2}{12}+\frac{e^{-y^2/3}}{x^4}-\frac{e^{y^2/6}}{x^2},
\label{pridscosfb}
\end{eqnarray}

\begin{eqnarray}
8\pi P_\bot/\alpha^2=-\frac{1}{6}-\frac{y^2}{12}-\frac{e^{-y^2/3}}{x^4},
\label{ptdscosfb}
\end{eqnarray}
and from (\ref{14a}) and (\ref{lum}) we obtain 
\begin{equation}
L_\Sigma=-\frac{ye^{-y^2/3}}{2},
\label{lums}
\end{equation}
and
\begin{eqnarray}
L_\infty=L_\Sigma e^{y^2/2}\left(1-\frac{ye^{y^2/6}}{6}\right)^2,
\label{lumsin} 
\end{eqnarray}
whereas the expressions for the complexity factor and the redshift read
\begin{equation}
Y_{TF}/\alpha^2=-\frac{1}{2}\left(\frac{y^2}{6}+1\right),
\label{bhf}
\end{equation}
and 
\begin{equation}
z_\Sigma=\frac{6 e^{y^2/4}}{6-y e^{y^2/6}}-1.
\label{zdis}
\end{equation}
Finally, the expression for $\tilde T$ becomes
\begin{eqnarray}
\tilde T=\frac{\ln x}{8\pi}\left[y-\tau^\ast \left(\frac{y^2}{3}-1\right)\right]+\tilde T_0(y),
\label{tdis}
\end{eqnarray}
or
\begin{eqnarray}
T^\ast=\frac{\ln x}{8\pi}\left[y-\tau^\ast \left(\frac{y^2}{3}-1\right)\right],
\label{tdisb}
\end{eqnarray}

where $T^\ast=\tilde T-\tilde T_\Sigma$. This choice of variable ($T^\ast$) allows us to skip the adoption of  a specific value for $\tilde T_\Sigma$.

Two comments are in order at this point
\begin{itemize}
\item As is follows from (\ref{5b1}) and (\ref{nd12}) the shear of our model is necessarily  non-vanishing.
\item The expansion scalar of our model vanishes, and we recall that  the expansion-free condition necessarily implies the appearance of a cavity
around the center \cite{exf}. Accordingly the fact that  the physical variables exhibit a singularity at $r=0$, does not represent a problem for the interpretation of the model. In other words the fluid under consideration is restricted to values of $x$ within the interval $[x_{min}>0, 1]$, where $x_{min}$ is the value of $x$ corresponding to the boundary of the cavity.
\end{itemize}

To highlight the effects derived from the appearance of the TKV on the evolution of the fluid we have plotted $Y_{TF}/\alpha^2$, $z_\Sigma$ and $L_\Sigma$, in Figure 1.  The dimensionless  physical variables $8\pi \mu/\alpha^2, 8\pi P_r/\alpha^2, 8\pi P_\bot/\alpha^2$ are plotted in Figure 2, whereas   $ 8\pi (P_r-P_\bot)/\alpha^2, 8\pi q/\alpha^2$ and $\alpha R_\Sigma$ are depicted in Figure 3. Finally in Figure 4 we plotted the variable $T^\ast$ for two different values of $\tau^\ast$ ($1$ and $0$).

All these results  will be discussed next.

\section{Discussion}
The problem of gravitational collapse is one of the most important problems  that astrophysicists %MDPI EE: Please check intended meaning has been retained. Ok
 have to face~\cite{20n,21n,22n}. For~decades, many researchers %MDPI EE: Please check intended meaning has been retained.Ok
  have devoted their work to elucidating some of the many questions raised by this issue. Numerical and analytical methods have been applied in such endeavors, each of which present advantages and~disadvantages.

In this work, we have adopted an analytical approach based on the assumption that at some point of the evolution, the~system momentarily admits a~TKV. 

Our main purpose was to exhibit the effects of~an instantaneous appearance of a TKV on~the evolution of relativistic self-gravitating fluids. %MDPI EE: Please check intended meaning.  has been retained.Ok
 This was achieved by means of  the concept of asymmetry factor introduced in~\cite{1}, which allows us to turn on (or off) a TKV at any given point of the~evolution. 

Our  model was assumed to be dissipative. Further specifications of this model are related to the specific choice of the two functions of $r$, up~to which the general model is defined, and~the choice of the form for the asymmetry factor $H_{TF}$; such choices obey to a simplicity~criterium. 

Under the conditions above, we obtain the model described by (\ref{muidscosfa})--(\ref{tdisb}). It represents an expanding fluid sphere for $y<0$ reaching a maximum at $y=0$ (the moment of the appearance of the TKV), contracting from there on, as~shown in the graphic of $\alpha R_\Sigma$ in Figure 3.

Since Darmois conditions are not satisfied on the boundary surface of the fluid distribution ($P_{r\Sigma}\neq q_\Sigma$), we have to assume that the boundary is a  thin shell. 
At this point, the following comments are in~order:
\begin{itemize}
\item The presented model was not conceived to describe any specific physical  scenario, but~just to bring out the effects produced by the sudden appearance of a TKV. Therefore, neither the presence of a thin shell, nor the violation of some energy conditions, are, at~this point, relevant for the discussion. All this having being said, it should be clear that the appearance of a thin shell implies that,  according to the Israel approach~\cite{isjc},   the~boundary is endowed with some specific material properties (surface energy-density and tension) whose role in the analysis of our model has not been taken into account. It goes without saying that in a model intended to describe a specific astrophysical object, the~dynamics of the shell might be a relevant factor and its role and the nature of its physical origin should be included in the discussion. A~similar  comment applies to any eventual violation of energy conditions.
\item As mentioned before, the choice (\ref{nref}) of the asymmetry factor %MDPI EE: Please check intended meaning has been retained. Ok.
 is justified by its simplicity. However, if we assume that $H(t)$ is a well-behaving function, we may expand it in a power series of $t-t_0$, implying that, at~least for~values of $t$ not far from $t_0$, the above choice is likely to describe the evolution quite well. Notwithstanding this, it is obvious that  $H(t)$ is a degree of freedom of the system, and~different choices may be invoked to describe different scenarios of physical interest. Thus, for example, one might propose a Gaussian  function, which would make the symmetry appear and then disappear smoothly (we thank one of the referees for this suggestion).
\item Our model was obtained under the assumption (\ref{deff}) for the functions $h(r)$ and $g(r)$, which as mentioned before, was adopted for its simplicity, and~leads to a particularly simple form of the metric functions in addition to~violations of some energy conditions as well as the appearance of thin shells. %MDPI EE: Please check intended meaning has been retained. Ok
 Although~such a choice allows us to bring out the effects of the sudden appearance (or disappearance)  of the TKV,  it goes without saying that for the treatment of any specific astrophysical scenario, the~choice of these functions must be dictated by the very nature of the problem under consideration.
\end{itemize}

 Some  remarkable  characteristics of this model, illustrated by Figure 1,Figure 2, Figure 3, Figure 4, deserve to be~emphasized:
\begin{itemize}
\item The complexity factor presents a local minimum (of its  absolute value) at $y=0$. The~existence of a stationary value of the complexity factor at $y=0$ (minimum or maximum) also appears in the case of a conformal Killing vector (CKV) analyzed in~\cite{1}, suggesting that a possible model independent link exists between complexity and symmetry.
\item The anisotropic factor presents a maximum at $y=0$. This is also observed in the case of the CKV, suggesting a model-independent link between the anisotropy factor \linebreak and symmetry.
\item There is a discontinuity in the surface redshift at around $y\approx 2.36$. Furthermore, the sign of $z_{\Sigma}$ changes %MDPI EE: Please check intended meaning has been retained. Ok
 along the evolution (from redshift to blueshift). This very peculiar behavior of the redshift is probably highly model-dependent. Notwithstanding this, it suggests the possibility of  observational support for the approach presented~here.

\item Particular attention should be given to the behavior of $T^\ast$ depicted in Figure~\ref{Figure 4}. In~the upper graphic, $T^\ast$ is plotted for $\tau^\ast=1$, and it was found that the same behavior (qualitatively) is observed for any value of $\tau^\ast \neq0$; namely,  $T^\ast$  decreases as $y$ approaches $0$, increasing afterward. However, if we assume that the relaxation time vanishes, then as~depicted in the corresponding graphic,  $T^\ast$ monotonically decreases after passing through $y=0$ (notice that, unlike the ``physical'' temperature $T$, the~variable $T^\ast$ may be negative). In~other words, the increase in  $T^\ast$ for $y>0$ is a relaxational effect that is expected to be  observed when the value of the relaxation time is equal to, or~larger  than, the~characteristic  time of the evolution. 
Thus, in~this specific model, we are able to detect the appearance of a thermal effect directly related to the relaxation time and the vanishing of the asymmetry~factor.  %MDPI EE: Please check intended meaning has been retained. Ok
{Before} %MDPI: Please confirm whether this part need to move to a new item.No, it does not.
ending this point, it is worth mentioning that we have resorted to the truncated version of the transport equation in order to calculate the temperature. We should recall that such an approximation is valid under certain physical conditions. However, we have not verified these conditions for our model %MDPI EE: Please check intended meaning has been retained.Ok.
 since, as~mentioned before, we do not intend to describe a specific astrophysical scenario. It is clear that %MDPI EE: Please check intended meaning has been retained. Ok.
  when dealing with this latter case, the~employed transport equation %MDPI EE: Please check intended meaning has been retained.Ok.
  must be properly adapted to the problem under consideration.
\end{itemize}

Thus, we may conclude that  the instantaneously appearance of a TKV produces important effects on the subsequent evolution of the system, some of which are directly related to observable variables (e.g., luminosity and the gravitational redshift). This points  to potential  applications of  the asymmetry factor approach in the study of self-gravitating systems. More specifically, it paves the way  to explain some observational astrophysical or cosmological data, which appears unintelligible based on the initial data alone, by~resorting to a general approach based on  the asymmetry factor. In~other words,  the~asymmetry factor approach seems to be particularly suitable for describing  self-gravitating fluid distributions exhibiting an unexpected behavior at some point of its evolution. However, for~a full justification of the method,  we should be able  to identify a physical reason for the vanishing of the asymmetry~factor.

We would like to conclude with the following comment. While it is true that specific imprints of the instantaneous appearance of a TKV are clearly model-dependent, the~very fact that such an appearance drastically changes the subsequent evolution of the system is not. In addition, for~some variables, there seems to exist %MDPI EE: Please check intended meaning has been retained. Ok.
 a persistent imprint for different models (e.g., the complexity factor and~the anisotropic factor).

\appendix
\section{Einstein equations}
 Einstein's field equations for the interior spacetime (\ref{1}) are given by
\begin{equation}
G_{\alpha\beta}=8\pi T_{\alpha\beta},
\label{2}
\end{equation}
and its non zero components
with (\ref{1}) and (\ref{3}) 
become
\begin{widetext}
\begin{eqnarray}
8\pi T_{00}=8\pi  \mu 
=\left(2\frac{\dot{B}}{B}+\frac{\dot{R}}{R}\right)\frac{\dot{R}}{R}
-\left(\frac{1}{B}\right)^2\left[2\frac{R^{\prime\prime}}{R}+\left(\frac{R^{\prime}}{R}\right)^2
-2\frac{B^{\prime}}{B}\frac{R^{\prime}}{R}-\left(\frac{B}{R}\right)^2\right],
\label{12b} \\
8\pi T_{01}=-8\pi qB
=-2\left(\frac{{\dot R}^{\prime}}{R}
-\frac{\dot B}{B}\frac{R^{\prime}}{R}\right),
\label{13} \\
8\pi T_{11}=8\pi P_r B^2
=-B^2\left[2\frac{\ddot{R}}{R}+\left(\frac{\dot{R}}{R}\right)^2\right]
+\left(\frac{R^{\prime}}{R}\right)^2-\left(\frac{B}{R}\right)^2,
\label{14} \\
8\pi T_{22}=\frac{8\pi}{\sin^2\theta}T_{33}=8\pi P_{\perp}R^2
=-R^2 \left[\frac{\ddot{B}}{B}+\frac{\ddot{R}}{R}
+\frac{\dot{B}}{B}\frac{\dot{R}}{R}\right]\nonumber \\
+\left(\frac{R}{B}\right)^2 \left[\frac{R^{\prime\prime}}{R}-\frac{B^{\prime}}{B}\frac{R^{\prime}}{R}\right].\label{15}
\end{eqnarray}
\end{widetext}

The component (\ref{13}) can be rewritten with (\ref{5c1}) and
(\ref{5b1}) as
\begin{equation}
4\pi qB=\frac{1}{3}(\Theta-\sigma)^{\prime}
-\sigma\frac{R^{\prime}}{R}.\label{17a}
\end{equation}

\end{document}